\documentclass[12pt]{article}
\usepackage{amssymb,amsmath}
\usepackage{bm} 
\usepackage{booktabs} 
\usepackage{array}
\usepackage{latexsym}
\usepackage{graphicx}
\usepackage{color}
\usepackage{datetime}
\usepackage[nosort]{cite}
\usepackage{verbatim}
\usepackage{enumerate}
\usepackage{chngpage} 
\usepackage{mathrsfs}
\usepackage{euscript}
\usepackage{psfrag}

\usepackage{datetime}

\usepackage[nosort]{cite}
\usepackage{chngpage} 
\usepackage{setspace}
\usepackage{tensor}

\usepackage[dvipsnames]{xcolor}

\usepackage{tensor}

\usepackage{mciteplus}

\usepackage[colorlinks=true,      linkcolor=blue,      urlcolor=blue,      
            filecolor=blue,      citecolor=blue,       pdfstartview=FitH,     
						pdfpagemode=UseNone,      bookmarksopen=true]{hyperref}  
\usepackage[all]{hypcap}     

\definecolor{cardinal}{rgb}{0.6,0,0}
\definecolor{darkgreen}{rgb}{0,0.4,0}
\definecolor{golden}{rgb}{0.92, 0.7, 0}
\definecolor{midnight}{rgb}{0, 0, 0.5}
\definecolor{darkblue}{rgb}{0, 0, 0.7}
\definecolor{purple}{rgb}{0.5, 0, 0.5}

\numberwithin{equation}{section}

\begin{document}

\begin{titlepage}

\bigskip\bigskip\bigskip
\bigskip

  \centerline{\LARGE \bf On the supersymmetries of branes with fluxes}

\medskip

\begin{center}

\vspace{14mm}

{\large
\textsc{Iosif Bena and Rapha\"el Dulac}}
\vspace{12mm}

\textit{ Institut de Physique Th\'eorique, \\  
Universit\'e Paris-Saclay, CNRS, CEA,\\
	Orme des Merisiers, Gif-sur-Yvette, 91191 CEDEX, France  \\} 
\medskip

\vspace{4mm} 
%

{\footnotesize\upshape\ttfamily  iosif.bena @ ipht.fr, raphael.dulac @ ens.fr } 
\vspace{13mm}
 

\end{center}

\centerline{Abstract}
\bigskip
The Dirac-Born-Infled action that describes the dynamics of D branes also allows one to compute the supersymmetries they preserve using the Kappa-symmetry projector. The ``Lagrangian'' expression of this projector depends on the velocity and electric fields of the branes, but not on the corresponding conserved charges. One can also construct the projector in a ``Hamiltonian'' approach, by multiplying the conserved string, brane and momentum charges with the corresponding gamma matrix involutions, adding them to the mass of the brane multiplied by the unit matrix, and normalizing the resulting expression. We show that these two procedures are equivalent.

\end{titlepage}
\tableofcontents

\label{sec:Intro}

\section{Introduction}
\label{sec:Intro}

There are several approaches to determining the amount of supersymmetry preserved by a particular system of branes. A single brane preserves 16 supersymmetries (half of the 32 of the vacuum), and adding another brane that is not attracted or repelled by the original brane results in a two-charge system that preserves 8 supersymmetries.  However, if the brane one adds is attracted by the original brane, they can form a bound state that still preserves 16 supersymmetries, but which differ from those of the original brane. For example, adding a D1 brane to a D3 branes forms a bound state that preserves 16 supercharges which are not those of the D3 branes.

There are two formalisms to determine the supersymmetries preserved by D-branes with fluxes and momentum. The first, which one can think of as a {\em Lagrangian} formalism, uses the DBI action of the D-brane, and the expression of the brane electric fields and brane velocities. This action has two fermionic symmetries: the global supersymmetry, coming from the spacetime supersymmetry of String Theory, and a world-sheet local fermionic symmetry. This second symmetry gives a projection equation for the 16 Killing spinors that are preserved locally on the brane worldvolume. One should note that the local existence of these Killing spinors does not give any indication on how much supersymmetry is preserved by the full configuration. The best example is the supertube \cite{Mateos:2001qs, Emparan:2001ux}, which preserves 16 supercharges at any location on its worldvolume, but only eight of these supercharges are common to all locations and hence are preserved by the full configuration. 

The second formalism, which we can think of as a {\em Hamiltonian} formalism, involves first computing the local charge densities of brane with fluxes, by taking functional derivatives of the brane Lagrangian with respect to the velocities and the worldvolume electric fields, $F_{0i}$. One then constructs a projector by adding to the unit matrix the ratios between these charge densities and the total mass density, multiplied by the gamma matrix corresponding to each charge. An example of the implementation of this procedure for the supertube can be found in \cite{Bena:2011uw}.

The purpose of this note is to show that the Lagrangian and the Hamiltonian procedures to determine the supersymmetry of a brane system are equivalent. This calculation could have been done more than twenty years ago, but it became more relevant because of the connection between supersymmetric brane configurations preserving sixteen supercharges locally and smooth supergravity solutions discussed in more detail in \cite{Bena:2022fzf, Bena:2022wpl,Bena:2023fjx}.

In Section \ref{section2} we review the construction of the ``Lagrangian''  $\kappa$-symmetry projector that annihilates the Killing spinors preserved by a brane with fluxes. In Section  \ref{section3} we review the construction of the ``Hamiltonian'' projector, which depends only on the conserved charges of the brane. We then proceed to show this equivalence for branes with increasingly complicated electromagnetic worldvolume fluxes: In Section  \ref{section4} we consider branes with purely electric and purely magnetic fluxes. In Section  \ref{section5} we show this equivalence for the supertube, which is the simples example where interacting magnetic and electric worldvolume fields are present. In Section \ref{section6} we show these two calculations give the same projector for the most general electric and magnetic fluxes on the brane.
\section{Supersymmetry projectors from the $\kappa$-symmetry}
\label{section2}

\subsection{Born-Infeld action and $\kappa$-symmetry}
The Born-Infeld action, first introduced as a non linear generalization of electromagnetism, and which appeared later as the low energy effective action of Dp-branes is: 
\begin{equation}
    S_{DBI}=-T_p\int \mathrm{d}^{p+1}\sigma e^{-\Phi} \sqrt{-\mathrm{det}\left(g_{ij}+ F_{ij}+B_{ij}\right)}\,,
\end{equation}
where $g_{ij}=g_{\mu\nu}\partial_iX^{\mu}\partial_jX^{\nu}$ and $B_{ij}$ are the pull-backs of the metric and spacetime NS-NS two-form on the worldvolume of the brane and $F$ is the field strength of the Abelian gauge field living on the worldvolume of the brane. This action has $10$ scalar degrees of freedom $X^{\mu}$, but only $9-p$ are physical one due to the $p+1$ diffeomorphism symmetry of this action, they give the embedding of the brane worldvolume in spacetime. These scalar fields,  $X^{I}$, are the transverse excitations of the brane. These, and the $p-1$ degrees of freedom of the gauge filed in $p+1$ dimensions give 8 bosonic degrees of freedom. The full bosonic part of the brane low-energy action is the sum of the Born-Infeld action and the Wess-Zumino action:
\begin{equation}
    S_{WZ}=-T_p \int \mathrm{d}^{p+1}\sigma e^{B_{ij}+ F_{ij}}\wedge \sum_{l}C^{(l)}\,.
\end{equation}
By abuse of language we use the name Born-Infeld action for the full low-energy action of branes. Our discussion of the $\kappa$-symmetry mainly follows \cite{
Becker:2006dvp, Simon:2011rw, Bergshoeff:1997kr}. 

The more natural way to understand the fermionic degrees of freedom living on the brane is to substitute the bosonic fields by their superspace counterparts. In addition to the reparameterization invariance, the Born-Infeld action is invariant under global supersymmetry. The SUSY parameter, $\varepsilon$, does not depend on the coordinates and has 32 components. This raises a paradox when one remembers that this action has 8 bosonic degrees of freedom while, the number of supersymmetries would suggest the presence of 16 fermionic degrees of freedom. This concern is solved thanks to a local fermionic symmetry, called $\kappa$-symmetry, which divides in half the number of preserved supersymmetries, and consequently the number of fermionic degrees of freedom. Under both supersymmetry and $\kappa$-symmetry, the fermions transform according to:
\begin{equation}
    \delta\Theta=\left(1-\Gamma\right)\kappa(\sigma)+\varepsilon\,,
    \label{eq:susy variation}
\end{equation}
with $\Gamma$ a traceless involution. For a purely bosonic background, $\Theta$ is trivial, hence $\delta\Theta=0$. Multiplying \eqref{eq:susy variation} by $1+\Gamma$, and requiring that the variation is still trivial leads to $(1+\Gamma)\varepsilon=0$. Consequently, the preserved supersymmetries are the nontrivial solutions of: 
\begin{equation}
    \left(1+\Gamma\right)\varepsilon_{\mathrm{unbr}}=0\,,
    \label{kappainv}
\end{equation}
where $\varepsilon_{\mathrm{unbr}}$ are the unbroken supersymmetries. As $\Gamma$ is a traceless involution, $\frac{1}{2}\left(1+\Gamma\right)$ is a projector with a kernel of dimension $16$. This results in a supersymmetric action that has the correct number of fermionic degrees of freedom.

\subsection{Using the  $\kappa$-symmetry to find preserved supersymmetries. }

Let us now review the construction of the $\kappa$-symmetry  projector,  $1+\Gamma_{\kappa}$, in a general background geometry with metric $h_{\mu \nu}$.

This section will be quite brief and mainly follows the formulation of \cite{ Bergshoeff:1997kr}. We will always use the generalized brane field strength $\mathcal{F}_{ij}=F_{ij}+B_{ij}$ where $F$ is the brane field strength and $B_{ij}$ the pullback on the brane of the spacetime NS-NS B-field. 

The pullback of the spacetime metric to the brane, $g_{ij}$, can be written as
\begin{equation}
    g_{ij}= {\partial x^{\mu}\over \partial x^i} {\partial x^{\nu}\over \partial x^j} h_{\mu \nu} \equiv 
    e_{i}^{a}e_{j}^{b}\eta_{ab}\,,
\end{equation}
where we have used the flat metric $\eta_{ab}$ to define the induced vielbeins $e_{i}^{a}$. This allows us to define the induced gamma matrices: 
\begin{equation}
    \gamma_{i}\equiv e_{i a}\Gamma^{a}\,,
\end{equation}
where $\Gamma^{a}$ are the flat space-time ten-dimensional gamma matrices.

The  full $\kappa$-symmetry  involution in \eqref{kappainv} is:

\begin{equation}\label{Kappa sym gamma}
    \Gamma=\frac{\sqrt{g}}{\sqrt{g+F}}\sum_{n}\frac{1}{2^nn!}F_{i_1k_1} \dots F_{i_{n}k_{n}}\gamma^{i_1 k_1 ... i_{n}k_{n}}J_{(p)}^{(n)}\,,
\end{equation}
with \begin{align}
   &J_{(p)}^{(n)}=\sigma_3^{n+\frac{p}{2}}i\sigma_2\Gamma_{(0)}\hspace{23.5mm}\text{Type IIA }\\
   &J_{(p)}^{(n)}=(-1)^n\sigma_3^{n+\frac{p-3}{2}}i\sigma_2 \Gamma_{(0)} \hspace{10mm}\text{Type IIB }\,
\end{align}
where $\Gamma_{(0)}=\frac{1}{(p+1)! \sqrt{g}}\epsilon^{i_0...i_p}\gamma_{i_0...i_p}$.
By construction $\Gamma^{2}=1$ and $\mathrm{Tr}(\Gamma)=0$, this confirms the fact that a brane breaks half of the supersymmetries of the vacuum.  In the rest of the paper we will refer to the Born-Infeld $\kappa$-symmetry formalism to compute the supersymmetries preserved by a brane as the \emph{Lagrangian} formalism.

\section{Supersymmetries in the Hamiltonian formalism}\label{Projector formalism for branes}

\label{section3}

Given a system of branes with various induced lower-brane charges, one can also determine the supersymmetries it preserves by calculating first all the conserved charges. For the D-branes dissolved in D-branes, these charge densities are given by the  magnetic components of the worldvolume field strength. However, if the branes have a null wave, or carry fundamental-string charges, the corresponding conserved charges and momenta are obtained by differentiating the Lagrangian with respect to the time derivatives of the scalars, or of the gauge potential. 

This allows us to calculate the Hamiltonian of this system, which depends on the conserved charges and momenta. To construct a projector for the corresponding system, one then multiplies the ratio between every charge and the Hamiltonian with the corresponding involution and sums over all the brane charges \cite{Bena:2011uw, Bena:2022wpl,Eckardt:2023nmn}. Thus the projector is
\begin{equation}
    \Pi = {1\over 2} \left( 1 + \sum_I{Q_I \over {\cal H}} P_I\right)\,,
    \label{proj-const}
\end{equation}
where the gamma matrix involution corresponding to every type of brane is:
\begin{equation}\nonumber
	\begin{array}{l|l}
		P_{\rm P}=\Gamma^{01} & P_{\rm  F1}=\Gamma^{01}\sigma_3\\ 
		P_{\rm D0}=\Gamma^{0}i\sigma_2 & P_{\rm D1}=\Gamma^{01}\sigma_1\\
		P_{\rm D2} =\Gamma^{012}\sigma_1 & P_{\rm D3}=\Gamma^{0123}i\sigma_2\\
		P_{\rm D4}=\Gamma^{01234}i\sigma_2 &P_{\rm  D5}=\Gamma^{012345}\sigma_1\\
		P_{\rm D6}=\Gamma^{0123456}\sigma_1&P_{\rm D7}=\Gamma^{01234567}i\sigma_2 \,.
	\end{array}
\end{equation}

Note that the fact that $\Pi^2=\Pi$ is equivalent to the mass-charge relation of the object we build. Furthermore, since  $\mathrm{Tr}(\Pi)=16$, for fixed $Q_I$ and $\cal H$, the system has 16 supersymmetries.

\subsection*{Global VS local SUSY}

It is important to stress that  $Q_I$ and ${\cal H}$ are in general local charge and energy densities, which can vary along the worldvolume of the brane. Hence, the projector \eqref{proj-const} should be more properly written as
\begin{equation}
    \Pi (\vec{x}) = {1\over 2} \left( 1 + \sum_I{Q_I (\vec{x}) \over {\cal H} (\vec{x})} P_I\right)\,.
    \label{proj-const2}
\end{equation}

At every value of $\vec x$ this projector preserves 16 supersymmetries, but these need not be the same. If we can find a constant non trivial $\varepsilon_0$ such that that $\Pi(\vec{x})\varepsilon_0=0$, then some of these 16 supersymmetries do not change as one moves along the brane, and hence the system preserves some global supersymmetries. When this happens, the projector can be written as
\begin{equation}
    \Pi(\vec{x})=f_1(\vec{x})\Pi_1+\ldots f_n(\vec{x})\Pi_p\,, \nonumber
\end{equation}
where $f_i(\vec x)$ can be nontrivial $\Gamma$-matrix valued functions, but the $\Pi_p$ are a set of commuting projectors that do not depend on $\vec{x}$. The brane configuration is therefore BPS and preserves $\frac{32}{2^{p}}$ supercharges. 


Thanks to this formalism, we can investigate the possible supersymmetries preserved by brane configurations, even before ascertaining whether these configurations  satisfy \emph{a priori} any equation of motion. The goal of this article is to show that, even off-shell, the  \emph{Hamiltonian} formalism presented above matches the \emph{Lagrangian} one.

\section{Equivalence of approaches in pure electric or magnetic solutions}
 \label{section4}
 
 To illustrate our method we first show that the two approaches give the same results for two simple brane bound states: a Dp brane with dissolved D(p$\rm-2$)  brane charge (or equivalently a constant worldvolume  magnetic field, $F_{12}$) and  a Dp-brane ($p\ge1$) with dissolved fundamental-string charge (or equivalently a constant worldvolume electric field $F_{01}$). 
 In the remaining sections we will consider more general branes with both electric and magnetic fields.
 
\subsection{ D(p$\rm-\mathbf{2}$)  branes within Dp-branes}

A Dp brane with a constant magnetic field (which we can take without loss of generality to be  $F_{12}$ has a nontrivial coupling with an electric  $C^{p-1}$ RR spacetime gauge field
\begin{equation}
    S_{WZ}=-\frac{1}{l_s^{p+1}(2\pi)^p}\int \mathrm{d}^{p+1}\sigma F_{12}\wedge C_{03...p}\,,
\end{equation}
and hence has a nontrivial D(p$\rm-2$)  charge.
The relation between the \emph{Lagrangian} and \emph{Hamiltonian} formulation of the supersymmetries is expected to be trivial as there is no velocity. 

The $\kappa$-symmetry projector is obtained by directly plugging in the magnetic field in a \eqref{Kappa sym gamma} and leads to:
\begin{equation}
    \Gamma=\frac{1}{\sqrt{1+F_{12}^2}}\left(1+F_{12}\Gamma^{12}\sigma_3\right)i\sigma_2\sigma_{3}^{p+1}\Gamma^{0...p}\,,
\end{equation}
which leads to a projection equation:
\begin{equation}
   \left(1+\frac{\Gamma^{0...p}i\sigma_2\sigma_{3}^{p+1}}{\sqrt{1+F_{12}^2}}+\frac{F_{12}}{\sqrt{1+F_{12}^2}}\Gamma^{03...p}i\sigma_2\sigma_{3}^{p}\right)\varepsilon=0\,.
\end{equation}

To compare this projector obtained in the Lagrangian formalism to the projector obtained in the Hamiltonian formalism, we need to compute the charges of this system. As there is no dynamical fields, the Hamiltonian is 
\begin{equation}
    \mathcal{H}=T_p\sqrt{1+F_{12}^2}\,,
\end{equation}
and the D(p$\rm-2$)  brane charge is $F_{12}$. We can therefore identify:
\begin{align}
    &\alpha_{Dp}\equiv {Q_p \over \cal H} =\frac{1}{\sqrt{1+F_{12}^2}}\\
    &\alpha_{D(p-2)}\equiv {Q_{p-2} \over \cal H} =\frac{F_{12}}{\sqrt{1+F_{12}^2}}\,.
\end{align}
As a consequence, from the Hamiltonian formalism we obtain the projection condition:
\begin{equation}
    \left(1+\alpha_p \Gamma^{0...p}i\sigma_2\sigma_{3}^{p+1}+\alpha_{p-2}\Gamma^{03...p}i\sigma_2\sigma_{3}^{p}\right)\varepsilon=0\,,
\end{equation}
which upon substituting the values of $\alpha_p, \alpha_{p-2}$ becomes:
\begin{equation}
    \left(1+\frac{\Gamma^{0...p}i\sigma_2\sigma_{3}^{p+1}}{\sqrt{1+F_{12}^2}}+\frac{F_{12}}{\sqrt{1+F_{12}^2}}\Gamma^{03...p}i\sigma_2\sigma_{3}^{p}\right)\varepsilon=0\,,
\end{equation}
As expected, this is exactly the same projector as that obtained in the Lagrangian $\kappa$-symmetry formulation. This result is rather simple but it is useful to illustrate the general philosophy.

\subsection{Fundamental strings dissolved in D-branes}
\label{F1inside}
A second example is a brane with a nontrivial electric field (which we can take without loss of generality to correspond to $F_{01}$). This brane has an extra charge, corresponding to fundamental strings along $1$ diluted into it. Since $F_{01}$ should be properly thought of as a velocity (as it becomes upon T-duality), the relation between the Lagrangian and the Hamiltonian formalism is slightly more involved. The $\kappa$-symmetry involution is:
\begin{equation}
    \Gamma=\frac{1}{\sqrt{1-F_{01}^2}}\left(1+F_{01}\Gamma^{01}\sigma_3\right)i\sigma_2\sigma_{3}^{p+1}\Gamma^{0...p}\,,
\end{equation}
so the projection equation is:
\begin{equation}
    \left(1+\frac{\Gamma^{0...p}i\sigma_2\sigma_{3}^{p+1}}{\sqrt{1-F_{01}^2}}+\frac{F_{01}\Gamma^{01}}{\sqrt{1-F_{01}^2}}\Gamma^{0123...p}i\sigma_2\sigma_{3}^{p}\right)\varepsilon=0 \,.
\end{equation}
It is possible to factorize by this by the Dp-brane involution $\Gamma^{0...p}i\sigma_2\sigma_{3}^{p+1}$ to obtain
\begin{equation}\label{fund string proj}
    \left(\Gamma^{0...p}i\sigma_2\sigma_{3}^{p+1}+\frac{1}{\sqrt{1-F_{01}^2}}-\frac{F_{01}}{\sqrt{1-F_{01}^2}}\Gamma^{01}\sigma_3\right)\varepsilon=0\,.
\end{equation}

On the Hamiltonian side, our solution has a fundamental string charge along the direction $1$, as well as charge corresponding to Dp-branes along the directions $1...p$. Hence,  projection equation is:
    \begin{equation}
    \left(1+\alpha_p\Gamma^{0...p}i\sigma_2\sigma_3^{p+1}+\alpha_F\Gamma^{01}\sigma_3\right)\varepsilon=0\,.
\end{equation}

To determine the $\alpha$'s we need to compute the conserved charge corresponding to the F1 strings, $\Pi$, as well as the Hamiltonian. 
\begin{align}
    &Q_{Dp}=T_p\\
    &Q_{F1}\equiv T_p \Pi= {\partial L \over \partial F_{01}} =\frac{T_{p}F_{01}}{\sqrt{1-F_{01}^2}}\\
    &H=\Pi F_{01} - L = T_{p} \sqrt{1+\Pi^2}\,.
\end{align}
This leads to a projection equation:
\begin{equation}
    \left(1+\frac{1}{\sqrt{1+\Pi^2}}\Gamma^{0...p}i\sigma_2\sigma_3^{p+1}+\frac{ \Pi}{\sqrt{1+\Pi^2}}\Gamma^{01}\sigma_3\right)\varepsilon=0\,.
    \label{fund-Ham}
\end{equation}
It is straightforward to see that by expressing $\Pi$ in terms of $F_{01}$ we obtain the projection equation \eqref{fund string proj} if we divide by the Hamiltonian density.

Note that in these equations we have not assumed anything about the form of $F_{01}$ or about the shape of the brane. The equivalence of the two formulation is a {\em local} one, and  works even when the worldvolume fields are arbitrary functions of the worldvolume coordinates. Note that this equivalence says nothing about how many global solutions equations \eqref{fund string proj} or equivalently \eqref{fund-Ham} have. 

\section{The supertube}

\label{section5}

We will now prove the equivalence of those two formalisms for branes with both worldvolume electric and magnetic fields, which are obtained when zooming in on the supertube.

Let us consider a brane with both an electric and a magnetic field, which have a leg in common: a Dp-brane ($p\ge2$) with electric field $F_{01}$ and magnetic field $F_{12}$. The Born-Infeld action is:
\begin{equation}
    S_{BI}=-T_p\int \sqrt{1-F_{01}^2+F_{12}^2}\,,
\end{equation}
and from now on we will work in units where $T_p=1$. The $\kappa$-symmetry involution is:
\begin{equation}
    \Gamma=\frac{1}{\sqrt{1-F_{01}^2+F_{12}^2}}\left(1+F_{01}\Gamma^{01}\sigma_3+F_{12}\Gamma^{12}\sigma_3\right)i\sigma_2\sigma_{3}^{p+1}\Gamma^{0...p},
\end{equation}
and the projection equation multiplied with the involution: $\left(1+F_{12}\Gamma^{12}\sigma_3\right)\Gamma^{0\ldots p}i\sigma_2\sigma_{3}^{p+1}$ becomes:
\begin{equation}
    \label{eq: Supertube kappa sym before hamiltonian}
    \left(\Gamma^{0\ldots p}i\sigma_2\sigma_{3}^{p}+F_{12}\Gamma^{03\ldots p}i\sigma_2\sigma_{3}^{p} \right)-\frac{F_{01}\Gamma^{01}\sigma_3}{\sqrt{1-F_{01}^2+F_{12}^2}}+\frac{F_{01}F_{12}\Gamma^{02}}{\sqrt{1-F_{01}^2+F_{12}^2}}\,.
\end{equation}
To identify the charges of the system, we need to compute both the conjugate momentum to $F_{01}$ and the Hamiltonian density:
\begin{align} 
    &Q_{\rm Dp}=T_p\\ \label{charges}
    &Q_{\rm F1}\equiv T_p \Pi^1 ={\partial L \over \partial F_{01}}  =\frac{T_p F_{01}}{\sqrt{1-F_{01}^2+F_{12}^2}}\\
    &Q_{\rm D(p-2)}=T_p F_{12}\\
    &\mathcal{H}=\frac{T_p(1+F_{12}^2)}{\sqrt{1+F_{12}^2-F_{01}^2}}=T_p\sqrt{(1+F_{12}^2)(1+(\Pi^1)^2)}\,.
\end{align}

The system also has a nontrivial momentum density along the direction $x_2$
\begin{equation}
  P_2 =\, F_{12} \Pi^1\,,
\end{equation}
coming from the $\left(E \times B\right)_{k}=\Pi^i F_{ik}$ Poynting vector on the brane worldvolume. For the supertube wrapping a closed curve, this momentum density contributes to the supertube angular momentum. It is also possible to understand this momentum density as a conserved charge which does not enter the Hamiltonian. This charge is nonzero because the presence of electric and magnetic fields induces a term in the Lagrangian that is linear in a velocity. 

Indeed, we can imagine giving the brane a tiny velocity, $v_2$, along the direction $x_2$. This velocity introduces a small $g^{\rm worldvolume}_{02} = g_{00}^{\rm spacetime} v_2 $. Since our supertube is in flat space, this conserved momentum conjugate to this velocity is:
\begin{equation}
T_p P_2=\left.\frac{\partial \mathcal{L}}{\partial v_2}\right|_{v_{2}=0} =\left.\frac{\partial \mathcal{L}}{\partial g_{02}}\right|_{g_{02}=0}= \frac{T_p F_{12} F_{01}}{\sqrt{1-F_{01}^2+F_{12}^2}}= T_p F_{12} \Pi^1\,.
  \label{Poynting}
\end{equation}

Thus, the supertube has both an F1 charge, $\Pi^1$ (the conjugate momentum of $F_{01}$), which appears quadratically in the Lagrangian and contributes to the Hamiltonian, as well as a momentum $P_2$ (the conjugate to $v_2$), which appears linearly in the Lagrangian and hence does not enter in the expression of the Hamiltonian.

In terms of these conserved charge densities, eq.\eqref{eq: Supertube kappa sym before hamiltonian} can be multiplied with $\frac{1}{\mathcal{H}}$ to give precisely the projector in the Hamiltonian formalism:
\begin{equation}
    \left(1-\frac{Q_{F1} }{\mathcal{H}}\Gamma^{01}\sigma_3+\frac{ P_2 \Gamma^{02}}{\mathcal{H}}+\frac{Q_{\rm D(p-2)}\Gamma^{03\dots p}i\sigma_2\sigma_3^{p}}{\mathcal{H}}+\frac{Q_{\rm Dp}\,\Gamma^{01\dots p}i\sigma_2\sigma_3^{p+1}}{\mathcal{H}}\right)\varepsilon=0\,. \label{supertube-proj}
\end{equation}

Since our analysis is local, it applies both to infinite Dp branes with lower brane charges, as well as to branes that have been rolled up along the direction of the momentum, into a  F1-D(p$\rm-2$) supertube with Dp dipole charge and a nontrivial angular momentum. As we explained above, whether or not the final configuration preserves any supersymmetry depends on the dimension of the common eigenspace of the projectors \eqref{supertube-proj} as one spans the worldvolume of the Dp brane. For two-charge supertubes, this dimension is 8 \cite{Mateos:2001qs}.

\section{General Brane-String Bound States}

\label{section6}

We are now ready to tackle the most general brane configuration, which has generic orthogonal momenta, as well generic electric and magnetic fields. There are two ways to approach this problem. 

The more formal way is to T-dualize this brane along all its worldvolume directions along which the electric and magnetic field strengths have legs. This results in a brane that is tilted very nontrivially and which has an orthogonal momentum. On can then T-dualize this brane along the direction of this orthogonal momentum, to recover the brane configuration in Subsection \ref{F1inside}. Since both the Hamiltonian and the Lagrangian determinations of the supersymmetries are local, and the equivalence of these formalisms is T-duality invariant, this constitutes a formal proof that the two formalisms agree.

However, this formal proof fails to reveal the rich physics and the subtle conserved charges that come into play when establishing the equivalence of the two formalisms.  We will therefore show how the equivalence can be established directly, by showing that the Lagrangian projector reproduces the Hamiltonian one for branes with arbitrary electric and magnetic fluxes. We will not consider branes that also have transverse momentum, because these branes can be T-dualized into branes with dissolved F1 charge, and the analysis is identical.

\subsection{Branes with general electric and magnetic fields}

Given a brane with worldvolume electric and magnetic fields, we can choose without loss of generality the worldvolume direction ``1'' to be the direction along the total electric field. Thus, only $F_{01}$ is non trivial, and all the other $F_{0j}$ vanish. We denote the remaining magnetic field strengths  $F_{ij}$. \footnote{We could do a further coordinate choice to label ``2'' the resulting direction of the magnetic field strength with a leg along ``1'', etc., but it is simpler to work with a general magnetic field.}

The expression under the square root of the Dp-brane Born-Infeld action is:
\begin{equation}
    \mathrm{Det}(g_{ij}+F_{ij})= \begin{vmatrix}
    -1 & -F_{01}& 0& \dots & 0\\
    F_{01} &1&-F_{12} &\dots &- F_{1p}\\
    0 & F_{12}& 1&\dots &- F_{2 p}\\
    \vdots & \vdots & &\ddots&\vdots\\
    0 & F_{1 p}&\dots&F_{p-1\, p} & 1\\
    \end{vmatrix}\,.
\end{equation}
By developing the determinant, we can define:
\begin{equation}
    \mathrm{Det}(g_{ij}+F_{ij})\!=\! -\!\begin{vmatrix}
   1&-F_{12} &\dots &- F_{1p}\\
    F_{12}& 1&\dots & F_{2 p}\\
     \vdots & &\ddots&\vdots\\
     F_{1 p}&\dots&F_{p-1\, p} & 1\\
    \end{vmatrix}+\!F_{01}^2 \begin{vmatrix}
     1&-F_{23}&\dots & -F_{2 p}\\
    F_{23}& 1&\dots&\vdots\\
    \vdots & \vdots &\ddots & \vdots\\
     F_{2 p}&\dots &\dots&  1\\
    \end{vmatrix}\equiv-A_{1\dots p}+F_{01}^2 A_{2\dots p}\,.
\end{equation}
 Therefore the full $\kappa$-symmetry projection equation can be written as:
 \begin{align}
   \!\!\!&\left( \!\! 1+\frac{1}{\sqrt{A_{1\dots p}\!-\!F_{01}^2\!A_{2\dots p}}}\!\!\left(\!1\!+F_{01}\Gamma^{01}\sigma_3 
      \!\!\!\!
   \sum_{\substack{n,\\i_j,k_j\neq 0,1}}
   \!\!\!\!
   \frac{1}{2^nn!}F_{i_1k_1} ... F_{i_{n}k_{n}}\Gamma^{i_1 k_1 ... i_{n}k_{n}}\sigma_3^{n}\!\!
   \right)\!\Gamma^{0\dots p}i\sigma_2\sigma_3^{p+1}\right)\varepsilon \nonumber \\
  \!\!\!   &+\frac{1}{\sqrt{A_{1\dots p}\!-\!F_{01}^2\! A_{2\dots p}}}\left(\left(\sum_{\substack{n\,>\,0,\\i_j,k_j\neq 0}}\frac{1}{2^nn!}F_{i_1k_1} ... F_{i_{n}k_{n}}\Gamma^{i_1 k_1 ... i_{n}k_{n}}\sigma_3^{n}\right)\Gamma^{0\dots p}i\sigma_2\sigma_3^{p+1}\!\!\right)\varepsilon=0
 \end{align}
 The term in the second line (with the $\frac{1}{\sqrt{A_{1\dots p}-F_{01}^2 A_{2\dots p}}}$ of the first line) would be the $\kappa$-symmetry involution (without the normalization factor)  if we did not have any electric field:
 
 \begin{equation}
\sqrt{A_{1\dots p}}\left.\Gamma_{\kappa}\right|_{F_{01}=0} =\sum_{\substack{n\\i_j,k_j\neq 0}}\frac{1}{2^nn!}F_{i_1k_1} ... F_{i_{n}k_{n}}\Gamma^{i_1 k_1 ... i_{n}k_{n}}\sigma_3^{n}\,,
\end{equation}

and this term squares to $A_{1\dots p}$. We multiply the projection equation by this term in order to obtain:
 \begin{align}
       \!\!\!\!\!&~~~0=\left(  \sum_{\substack{n,\\i_j,k_j\neq 0}}\frac{1}{2^nn!}F_{i_1k_1} ... F_{i_{n}k_{n}}\Gamma^{i_1 k_1 ... i_{n}k_{n}}\sigma_3^{n}\right)\Gamma^{0\dots p}i\sigma_2\sigma_3^{p+1}\varepsilon+\frac{A_{1\dots p}}{\sqrt{A_{1\dots p}\!-\!F_{01}^2\! A_{2\dots p}}}\varepsilon\,+\\
      \!\!\!\!\! &\left(
         \!\!\!\!\!\!\!\! 
     \sum_{\substack{n,\\
     ~~~~i_j,k_j\neq 0}}
    \!\!\!\!\!\!
     \frac{1}{2^nn!}F_{i_1k_1} ... F_{i_{n}k_{n}}\Gamma^{i_1 k_1 ... i_{n}k_{n}}\sigma_3^{n}\!\right)\!\!\frac{F_{01}\Gamma^{01}\sigma_3}{\sqrt{A_{1\dots p}\!-F_{01}^2 A_{2\dots p}}}\!\!\left(
     \!\!\!\!\!\!\!\!
     \sum_{\substack{n,\\
     ~~~~i_j,k_j\neq 0,1}}
     \!\!\!\!\!\!\!\!\frac{(-1)^{n+1}}{2^nn!}F_{i_1k_1} ... F_{i_{n}k_{n}}\Gamma^{i_1 k_1 ... i_{n}k_{n}}\sigma_3^{n}\!\!\right)\!\varepsilon \nonumber
 \end{align}
 At this point it is useful to compute the momentum conjugate to $F_{01}$ and the Hamiltonian density:
 \begin{equation}
     \Pi_{\rm F1_1}=\frac{F_{01}A_{2\dots p}}{\sqrt{A_{1\dots p}-F_{01}^2 A_{2\dots p}}}\,,\quad \mathcal{H}=\sqrt{\left(1+\Pi_{\rm F1_1}^2 A_{2\dots p}\right)\left(A_{1\dots p}\right)}
     \label{Ham-momentum}
 \end{equation}
 We then divide the projection equation by $\mathcal{H}$ everywhere, to obtain:
 \begin{align}
    \label{eq:General formula hamiltonian}
     &0= \left( 1+ \sum_{\substack{n,\\i_j,k_j\neq 0}}\frac{1}{2^nn!}\frac{F_{i_1k_1} ... F_{i_{n}k_{n}}}{\mathcal{H}}\Gamma^{i_1 k_1 ... i_{n}k_{n}}\sigma_3^{n}\Gamma^{0\dots p}i\sigma_2\sigma_3^{p+1}\right)\varepsilon \,+
    \\
     &\left(\sum_{\substack{n,\\i_j,k_j\neq 0}}\!\!\!\frac{1}{2^nn!}F_{i_1k_1} ... F_{i_{n}k_{n}}\Gamma^{i_1 k_1 ... i_{n}k_{n}}\sigma_3^{n}\right)\!\!\frac{\Pi_{\rm F1_1}\! \Gamma^{01}\sigma_3}{A_2\mathcal{H}}\!\!\left(\sum_{\substack{n,\\i_j,k_j\neq 0,\,1}}\!\!\!\!\frac{(-1)^{n+1}}{2^nn!}F_{i_1k_1} ... F_{i_{n}k_{n}}\Gamma^{i_1 k_1 ... i_{n}k_{n}}\sigma_3^{n}\!\right)\varepsilon\,.\nonumber 
 \end{align}
 
 We would like to claim that this formula is the brane supersymmetry projector in the Hamiltonian formalism. As one can read off from the Wess-Zumino action, the first line contains the Dp charge density and all the lower D-brane charge densities dissolved in our Dp brane, multiplied by their corresponding $\Gamma$-matrices and divided by the Hamiltonian density, as in \eqref{proj-const}.
 
The second line is then expected to contain a term associated to the F1 charge,  $\frac{\Pi_{\rm F1_1}}{\mathcal{H}}$ multiplied by the $\Gamma$-matrix involution corresponding to F1 strings extended along the direction ``1''. It is not hard to see that the term proportional to the identity matrix in 
 \begin{equation}
    \left(\sum_{\substack{n,\\i_j,k_j\neq 0,1}}\frac{1}{2^nn!}F_{i_1k_1} ... F_{i_{n}k_{n}}\Gamma^{i_1 k_1 ... i_{n}k_{n}}\sigma_3^{n}\right)\left(\sum_{\substack{n,\\i_j,k_j\neq 0}}\frac{(-1)^{n+1}}{2^nn!}F_{i_1k_1} ... F_{i_{n}k_{n}}\Gamma^{i_1 k_1 ... i_{n}k_{n}}\sigma_3^{n}\right)
\end{equation}
is exactly equal to $A_2$. Hence, the term in the projector proportional to the F1 charge is reproduced. 

The second line contains many other terms, whose interpretation is more subtle and will be illustrated in detail in the next subsection. There is a term that contains the product of the momentum conjugate of the electric field, $\Pi_{\rm F1_1}$, with a single magnetic field, $F_{1i}$. This describes the Poynting vector that gives rise to a nontrivial conserved momentum, $p_i$, as in \eqref{Poynting}. It is not hard to see that the combination of these fields is multiplied by the appropriate $\Gamma$-matrix involution to correspond to this momentum charge. 

Then there are more complicated terms, which contain multiple products of fields. All these terms correspond to either fundamental-string or momentum charges (depending on whether $(\sigma_3)^n=1\,\text{or}\,\sigma_3$). The conserved F1 charges come from the frame dragging of the magnetic fields $F_{i\,j}$ by the Poynting-vector-induced momenta $p_i$. The value of these conserved F1 charges can be found by turning on an infinitesimal $F_{0j}$ and computing the conjugate momenta: 
\begin{equation}
Q_{\rm F1_j}= \left. {\partial L \over \partial F_{0j}} \right|_{F_{0j}=0}\,.
\end{equation}

In turn, the combination of these new F1 charges with the existing magnetic fields gives rise to other Poynting vectors and therefore to momenta along other directions. Exactly like in \eqref{Poynting}, the values of these momenta can be computed by taking partial derivatives of the Lagrangian density with respect to an infinitesimal velocity along these directions:
\begin{equation}
{\rm P_j}=  \left. {\partial L \over \partial v_j} \right|_{v_{j}=0}= \left. {\partial L \over \partial g_{0j}} \right|_{g_{0j}=0}\,.
\end{equation}
These momenta, in turn, induce other conserved F1 charges, which in turn induce other conserved momenta, and so forth. All these conserved charges enter in the Hamiltonian expression of the supersymmetry projector and it is a straightforward exercise to show that they are exactly the terms that appear in the second line of the Lagrangian projector, \eqref{eq:General formula hamiltonian}. In the next subsection we will illustrate this explicitly.  

 \subsection{A illustrative example} 

 To illustrate the physics described in the previous subsection, it is best to focus on a brane with three types of field strength: $F_{01}\,,F_{12}$ and $F_{23}$. The $\kappa$-symmetry involution is
\begin{equation}
    \Gamma_{\kappa}=\frac{1}{\sqrt{(1-F_{01}^2)(1+F_{23}^2)+F_{12}^2}}\left(\left(1+F_{01}\Gamma^{01}\sigma_3\right)\left(1+F_{23}\Gamma^{23}\sigma_3\right)+F_{12}\Gamma^{12}\sigma_3\right)\Gamma^{01\dots p}i\sigma_2\sigma_3^{p+1}\,.
\end{equation}
The obvious F1 conserved charge, $\Pi$, is the conjugate momentum to $F_{01}$. Its formula and that of the Hamiltonian can be obtained straightforwardly from \eqref{Ham-momentum} (with $T_p=1$):
 \begin{align}
    \Pi=\,\frac{F_{01}\,(1+F_{23}^2)}{\sqrt{F_{12}^2+(1-F_{01}^2)(1+F_{23}^2)}}\,,\quad\mathcal{H}=\sqrt{\frac{(1+F_{12}^2+F_{23}^2)(1+F_{23}^2+\Pi^2)}{1+F_{23}^2}}
\end{align}
The Lagrangian projection equation, \eqref{eq:General formula hamiltonian}, becomes: 
\begin{align}
    \label{eq:3 field kappa sym}
    &\left(1+\frac{\Gamma^{01\dots p}i\sigma_2}{\mathcal{H}}+\frac{F_{12}\,\Gamma^{03\ldots p}i\sigma_2\sigma_3^{p}}{\mathcal{H}}+\frac{F_{23}\,\Gamma^{014\ldots p}i\sigma_2\sigma_3^{p}}{\mathcal{H}}\right)\varepsilon \nonumber \\
    &+\left(-\frac{\Pi\, \Gamma^{01}\sigma_3}{\mathcal{H}}\,+\,\frac{\frac{\Pi F_{12}}{1+F_{23}^2}\, \Gamma^{02}}{\mathcal{H}}\,-\,\frac{\frac{\Pi F_{12}F_{23}}{1+F_{23}^2}\, \Gamma^{03}\sigma_3}{\mathcal{H}}\right)\varepsilon=0\,.
\end{align}

We would like to show that this is same as the Hamiltonian projector \eqref{proj-const}. The first line represents the contribution to the projector from the D-branes charges, which are independent of the electric field. The first term in the second line is the contribution to the projection equation of the F1 charge along the direction ``1''. As we mentioned above, although this is not obvious from the general equation \eqref{eq:General formula hamiltonian}, the coefficient of this term is exactly the one we expect.

The coefficient of the second term, $\frac{\Pi F_{12}}{1+F_{12}^2}$, is the conserved momentum along the direction ``2'', and it comes from a Poynting vector generated by the interaction of the electric field along ``1'', whose conserved momentum conjugate is $\Pi=Q_{\rm F1_1}$ and the magnetic field $F_{12}$. This momentum does not enter the expression of the Hamiltonian. Its value is obtained by taking a partial derivative with respect to a vanishing velocity along ``2'':
\begin{equation}
  P_{2} = 
        \left.\frac{\partial \mathcal{L}}{\partial g_{02}}\right|_{g_{02}=0}=
    \frac{\Pi\, F_{12}}{1+F_{12}^2}\,.
\end{equation}

The third term corresponds to an F1 charge, along the direction ``3''. This charge can be thought of as coming from the frame dragging induced by the momentum $P_2$ and the magnetic field $F_{23}$. To calculate this charge we again turn on an infinitesimal $F_{03}$ term and compute:
\begin{equation}
Q_{\rm F1_3} =   \left.\frac{\partial \mathcal{L}}{\partial F_{03}}\right|_{F_{03}=0}=\frac{\Pi\,F_{12}\,F_{23}}{1+F_{23}^2}\,.
\end{equation}

Thus, all the terms in \eqref{eq:3 field kappa sym} have the correct structure that reproduces the Hamiltonian computation of the supersymmetry projector \eqref{proj-const}.

\vspace{1em}\noindent {\bf Acknowledgements:}  We would like to thank Yixuan Li and Masaki Shigemori for interesting discussions. The work of IB was supported in part by the ERC Grants {\em 787320 - QBH Structure} and {\em 772408 - Stringlandscape} and by the NSF grant PHY-2309135 to the Kavli Institute for Theoretical Physics (KITP).

\bibliographystyle{utphys}   
\bibliography{biblio.bib}

\end{document}